\documentclass[twocolumn]{emulateapj}

\usepackage{epsfig}
\usepackage{amsmath}
\usepackage{natbib}

\begin{document}

\title{Quasar Microlensing: when compact masses mimic smooth matter}
\author{Geraint F. Lewis \& Rodrigo Gil-Merino}
\email{gfl@physics.usyd.edu.au}
\email{rodrigo@physics.usyd.edu.au}
\affil{Institute of Astronomy, School of Physics, A28, 
University of Sydney, NSW 2006, Australia}

\begin{abstract}
The magnification  induced by gravitational  microlensing is sensitive
to the size  of a source relative to the  Einstein radius, the natural
microlensing  scale  length. This  paper  investigates  the effect  of
source size in the case  where the microlensing masses are distributed
with a bimodal  mass function, with solar mass  stars representing the
normal  stellar  masses,  and   smaller  masses  (down  to  $8.5\times
10^{-5}$M$_\odot$) representing a dark  matter component.  It is found
that there exists a critical regime where the dark matter is initially
seen as individual compact masses,  but with an increasing source size
the compact dark  matter acts as a smooth  mass component.  This study
reveals that  interpretation of microlensing  light curves, especially
claims of small mass dark matter lenses embedded in an overall stellar
population, must consider  the important influence of the  size of the
source.
\end{abstract}


\maketitle

\section{Introduction}
As the light from a distant source shines through a foreground galaxy,
the gravitational  lensing effect of  the individual stars  within the
galaxy can induce rapid fluctuations in the apparent brightness of the
source \citep{1986ApJ...301..503P}.  Since  the detection of the first
lensing  induced  variability, seen  in  the  quadruply imaged  quasar
Q2237+0305  \citep{1989AJ.....98.1989I,1991AJ....102...34C}, there has
been   intense   theoretical   study   of  this   {\it   gravitational
microlensing}, revealing that the degree of brightness fluctuations is
strongly  dependent   upon  the  size  of  the   source  being  lensed
\citep{1986A&A...166...36K,  1990ApJ...358L..33W, 1991AJ....102..864W}
and     the     mass    function     of     the    lensing     objects
\citep{1996MNRAS.283..225L,2001MNRAS.320...21W}, potentially providing
a useful probe of both. Further studies have scrutinized the influence
of     source    size     and     shape    \citep{1991A&A...250...62R,
1993A&A...278L...5R,     1997A&A...325..877R,     2003ApJ...594..449A,
2005ApJ...628..594M},   and  more   recent  programs   with  dedicated
monitoring  programs  \citep{2000ApJ...529...88W, 2002ApJ...572..729A,
2003MNRAS.346..415U}   have   finally   resulted   in   high   quality
microlensing  light   curves  that  can  be   compared  directly  with
theoretical models~\citep{2000MNRAS.318.1120W,2004ApJ...605...58K}.

\begin{figure*}
\centerline{ \psfig{figure=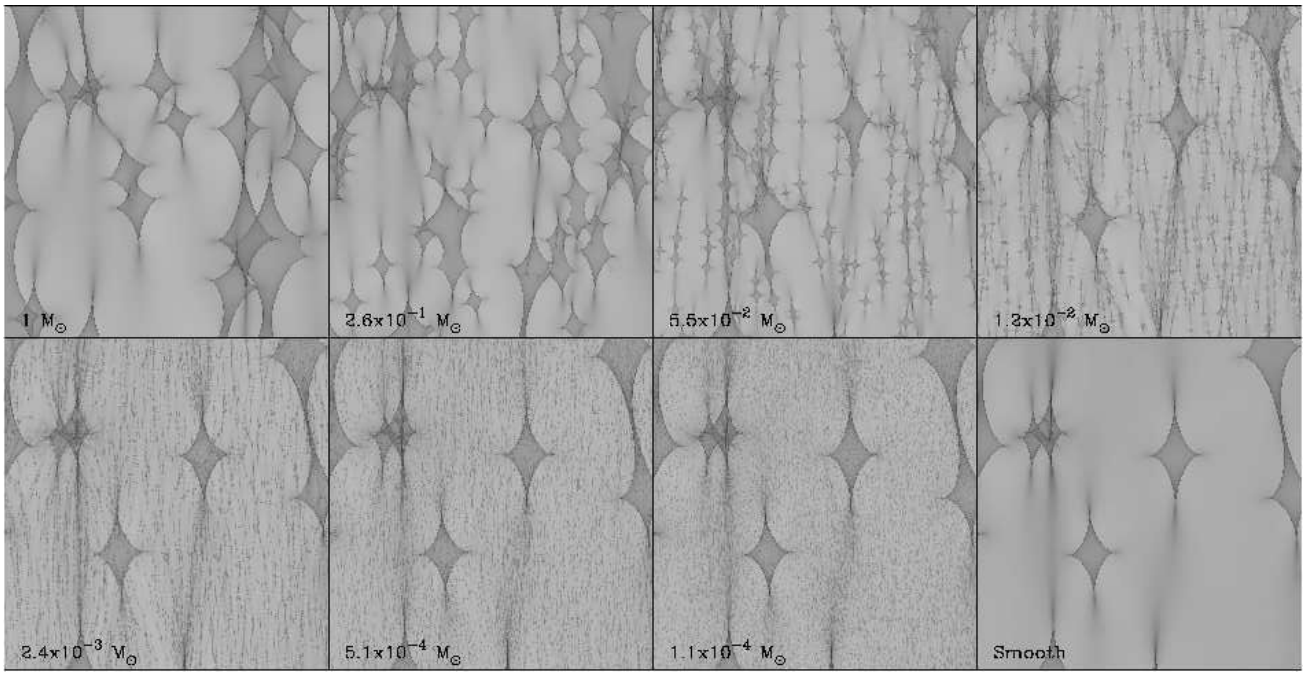,angle=270,width=6.8in}}
\caption[]{Examples of the magnifications maps utilized in this paper.
Each map is 12 Einstein radii on a side with a normalized surface mass
density  of  $\sigma=0.2$  and  global  shear  of  $\gamma=0.5$.   The
lower-right panel  presents the  case where half  of the  surface mass
density is in the form of solar  mass objects and the rest in the form
of smoothly distributed matter.  In the additional panels, this smooth
matter  component has  been replaced  with compact  mass  objects with
solar  and  then  sub-solar  masses;  the  individual  masses  of  the
sub-solar population are  noted in the lower left-hand  corner of each
panel.  Note  that in  each panel, the  solar mass stars  retain their
positions in the lensing plane.
\label{fig1}}
\end{figure*}

More  recently, gravitational  microlensing  has been  proposed as  an
explanation  of the  anomalous  flux ratios  seen  in multiply  imaged
quasars,  with  the  addition  of  smoothly  distributed  dark  matter
broadening  the width  of the  magnification  probability distribution
\citep{2002ApJ...580..685S}.  This  result appears somewhat surprising
given  a  conjecture  that  has  existed  for a  number  of  years  in
microlensing  studies,  namely  that  the  shape  of  a  magnification
probability distribution should be independent of the mass function of
the                         microlensing                        masses
\citep{1992ApJ...386...19W,1996MNRAS.283..225L};   the  smooth  matter
limit  is  approached  as  the  mass  of  the  individual  objects  is
decreased, but the surface mass density remains the same. Clearly, the
significant   differences  seen   in  the   magnification  probability
distributions with smooth and compact matter violates this conjecture,
a position recently confirmed by \citet{2004ApJ...613...77S}.

The fall  of the microlensing conjecture therefore  implies that there
is a fundamental difference in the action of compact masses as opposed
to smoothly distributed mass. But  can compact masses appear as smooth
matter to  a source? It  is well known  that increasing the size  of a
microlensing source  smooths out the  variability of the  light curve,
but does  this imply  that to  a large source,  a population  of small
masses  appears  as  smooth  matter?   While this  situation  was  not
examined  in  the work  of  \citet{2004ApJ...613...77S}, it  existence
means  that the breakdown  of the  microlensing conjecture  is further
blurred by the  size of the source under  consideration. Studying this
regime,    therefore,   is    the    goal   of    this   paper.     In
Section~\ref{approach},  the  numerical  approach is  outlined,  while
Section~\ref{analysis} discusses  the influence  of a range  of source
sizes and compares the resulting properties of the magnification maps.
In Sections~\ref{implications} and \ref{conclusions}, the implications
of this study and the conclusions are respectively presented.

\section{Approach}\label{approach}
In studying  the influence  of gravitational microlensing,  this study
employs     the     backwards     ray     tracing     algorithm     of
\citet{1986A&A...166...36K} and \citet{1992ApJ...386...19W}, whereby a
large number of rays are  fired through a field of microlensing masses
and collected to form a  magnification map over the source plane.  The
important length scale for  gravitational microlensing is the Einstein
radius in the source plane, given by
\[
\eta_o = \sqrt{ \frac{4 G M}{c^2} \frac{D_{os} D_{ol}}{D_{ls}} }
\]
where  $D_{ij}$ are  angular diameter  distances between  the observer
($o$),  lens ($l$) and  source ($s$)  respectively and  typically this
scale   is   $\eta_o\sim   0.04$pc  for   cosmological   circumstances
\citep[e.g.][]{1998ApJ...501..478L}. As  well as the  mass function of
the  microlensing  bodies,  other  parameters need  to  be  specified,
including the normalized surface  mass density ($\sigma$) which may be
composed of  compact and  smooth masses, and  a shear  term ($\gamma$)
which describes the  global asymmetric gravitational lensing influence
of  the overall  galactic/intracluster mass  distribution.   Given the
potential  parameter   space,  a  comprehensive   study  of  potential
microlensing configurations is  beyond the scope of this  paper.  As a
first step,  a representative  model was chosen  with a  total surface
mass density of $\sigma=0.2$ and global shear of of $\gamma=0.5$. Each
simulation covers  an area of $12^2$  Einstein radii for  a solar mass
star and consists of $2048^2$ pixels.

\begin{figure*}
\centerline{ \psfig{figure=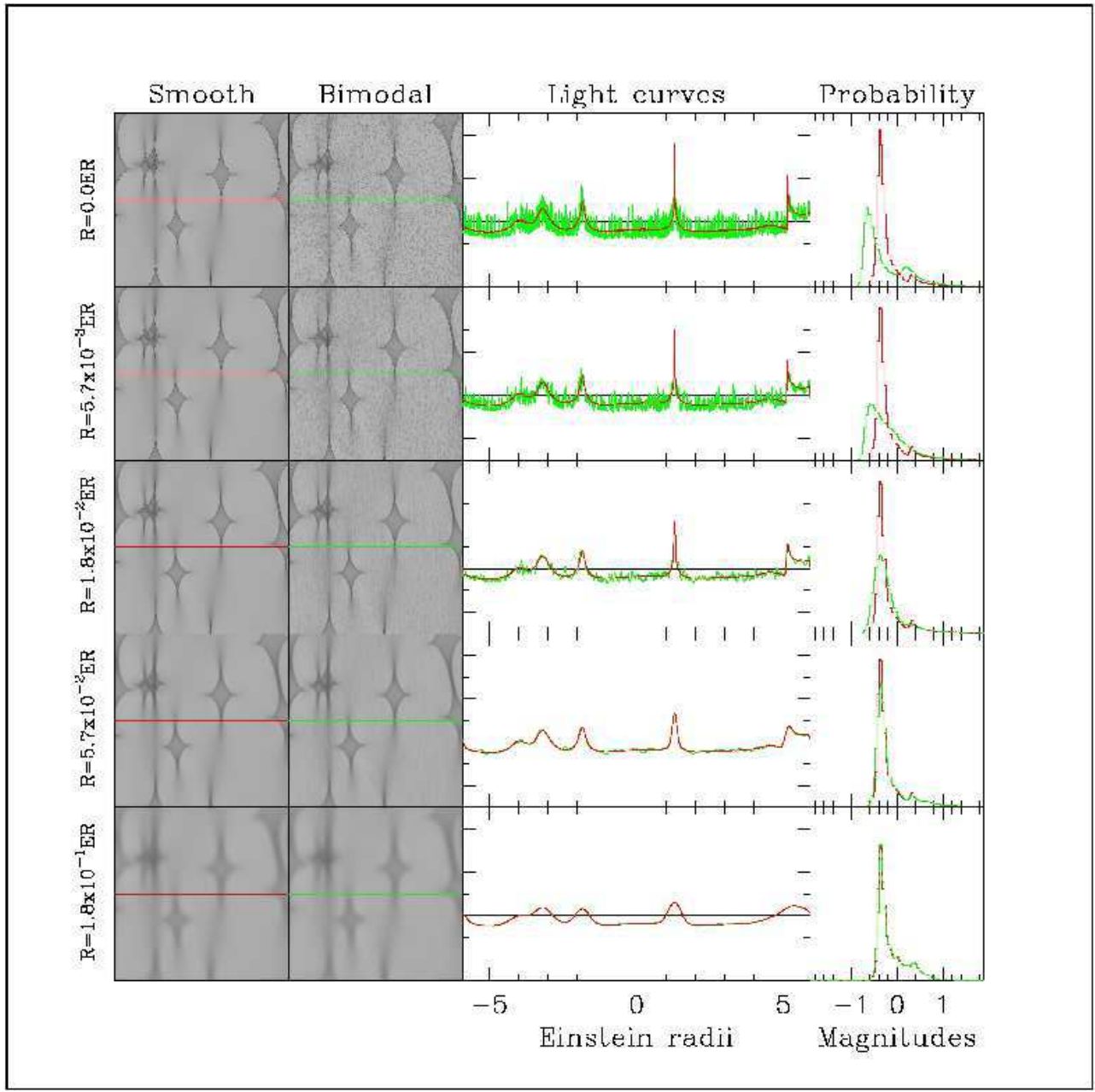,angle=270,width=6.7in}}
\caption[]{An example  of the procedure undertaken in  this paper. The
left-most  column contains a  magnification map  formed when  half the
surface mass density is in  solar mass objects, with the rest smoothly
distributed. The  centre panel presents the  same circumstance, except
the  smooth matter component  has been  replaced with  compact objects
each with  a mass of $8.5\times10^{-5}{\rm  M_\odot}$. Following these
are light curves for the  smooth matter (red) and compact mass (green)
cases, taken across the centre of  the maps (indicated by the lines on
the magnification maps).  Note that  the light curves are in magnitude
and  the  range on  the  y-axis  from +3  to  -5  magnitudes about  the
theoretically expected mean. This  scale has been omitted for clarity.
The   right  hand  column   presents  the   magnification  probability
distributions.  The  top maps represent  a single pixel  source, while
moving downwards the source size  is increased.  As it does, the small
scale caustics structure  induced by the smaller masses  is washed out
and the maps, light  curves and magnification distributions become the
same.  While the  top panel  is unconvolved,  the scale-radius  of the
Gaussian  kernel  in  the  subsequent panels  are  $5.9\times10^{-3}$,
$1.8\times10^{-2}$, $5.7\times10^{-2}$ and $1.8\times10^{-1}$ Einstein
radii respectively.
\label{fig2}}
\end{figure*}

A fiducial model was chosen in  which half the surface mass density is
in the  form of solar mass stars,  while the remaining mass  is in the
form  of  smoothly  distributed  matter.   In  addition  to  this,  43
additional simulations  were undertaken. Again,  half the mass  was in
the form of solar mass stars,  while the remainder is also in the form
of compact  objects, but with progressively smaller  masses, with each
subsequent dark matter  mass being 80\% of the  proceeding value.  The
limit   of  this  procedure   was  compact   dark  matter   masses  of
$8.5\times10^{-5}$M$_\odot$,   resulting   in   a  total   number   of
microlensing  masses in the  simulation exceeding  $4.2\times10^6$; it
was  found  that  this  represented  the  memory  limit  of  available
computational hardware, while  greatly increasing the calculation time
of  the   magnification  maps,  and  hence  no   smaller  masses  were
considered.  It is  important to note that the  locations of the solar
mass stars are identical in all simulations.

Figure~\ref{fig1} presents  examples of the results  of this approach,
with the upper-left panel  representing the magnification maps for the
case where  half the matter is in  the form of smooth  matter. For the
additional  panels in this  figure, this  smooth matter  component has
been replaced  with compact  objects, beginning with  $1$M$_\odot$ and
continuing down to $1.1\times10^{-4}$M$_\odot$. There is a significant
difference   between   the   caustic   structures  apparent   in   the
magnification  map for the  smooth matter  case and  that for  all the
lensing  bodies being  of 1M$_\odot$.   However,  as the  mass of  the
compact  objects replacing  the  smooth matter  component is  reduced,
there  is  a  very apparent  evolution  in  the  form of  the  caustic
structure; maps  appear to possess caustics structure  on two distinct
scales,   corresponding   to  the   two   microlensing  masses   under
consideration.  For  the magnification  map with the  smallest masses,
the large scale caustic structure  becomes very similar to that of the
case with  smooth matter. However, it  is also apparent  that there is
significant small  scale structure distributed throughout  the map. As
will  be seen in  the next  section, while  the magnification  maps in
smooth  matter and small  mass cases  appear similar,  the statistical
properties,  especially the  magnification  probability distributions,
differ markedly.

\section{Analysis}\label{analysis}
To  determine the  microlensing  properties of  a particular  extended
source, the magnification maps  must be convolved with the appropriate
surface brightness  distribution. For the purposes of  this study, the
source is assumed to possess  a Gaussian profile with a characteristic
radius. As  the source size is increased,  the resulting magnification
map and magnification probability distribution are compared with those
of the solar mass plus smooth matter component.

Figure~\ref{fig2} illustrates  this procedure  for the case  where the
compact   dark   matter    component   have   individual   masses   of
$8.5\times10^{-5}$M$_\odot$,    the   smallest    masses    that   are
computationally  viable at this  time.  The  upper panels  present the
unconvolved magnification maps for  the smooth dark matter (left-hand)
and compact dark  matter (right-hand)\footnote{It should be remembered
that these  maps, by  virtue of the  pixelated nature,  are implicitly
convolved with  a source on the  scale of a single  pixel.}.  The maps
possess  similar  large  scale  caustic  structure,  but  clearly  the
presence of small masses instead  of smooth dark matter has introduced
significant  substructure into the  caustic network.   Over-plotted on
these magnification  maps are the  resultant light curves of  a source
passing  horizontally across  the middle  of  the two  maps and  these
present  quite   dramatic  behaviour.    This  is  reflected   in  the
magnification probability distributions  which are markedly different.
Progressing down the page  these magnification maps are convolved with
larger  and  larger  sources.   As  the  source  size  increases,  the
substructure in the magnification  map with compact dark matter begins
to wash out. This is further reflected in the subsequent light curves,
with  the degree  of  variability decreasing.   Interestingly, as  the
source size is increased, the light curves for the compact dark matter
and correspondingly convolved smooth matter cases become very similar,
again a fact seen in the magnification probability distributions which
too become  identical.  With even larger  smoothing, the magnification
maps remain  identical and hence  the compact matter  will effectively
appear as a smooth dark matter component.

\begin{figure}
\centerline{ \psfig{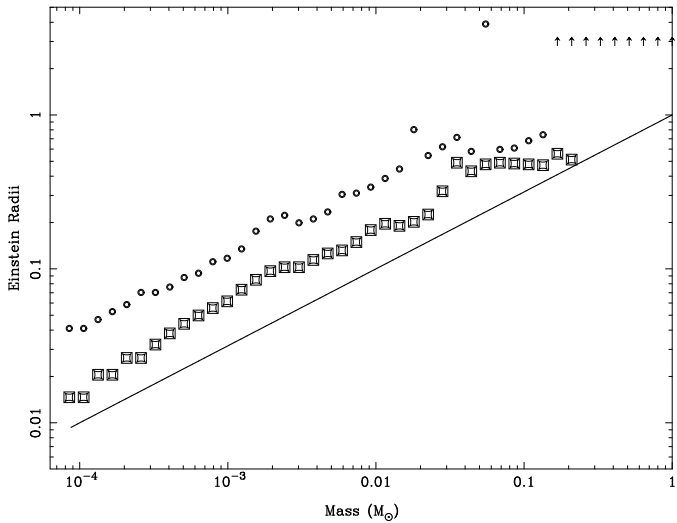}}
\caption[]{The solid  line in this figure denotes  the Einstein radius
for  the smaller  mass  microlensing simulations,  whereas the  points
present the radius  of the Gaussian kernel at  which the difference in
the binned,  cumulative magnification  patterns differed by  less than
$f=0.05$ (circles) and $f=0.10$  (squares); note the relation shows an
linear trend in this logarithmic space.  The discrepant points are due
to  ``noise''  in the  binned  distribution  preventing  the fit  from
reaching the fitting  criteria.  The arrows in this  figure denote the
magnification  maps  which did  not  achieve  this  criterion and  the
magnification distributions are too  different to allow an equivalence
between the two maps even with smoothing on the scales considered.
\label{fig3}}
\end{figure}

To examine this  further, the entire sample was  subjected to the same
convolution  procedure,  starting  with  small  Gaussian  sources  and
increasing  the   radius  of   the  kernel  until   the  magnification
probability distributions became  similar. In defining this threshold,
the binned cumulative  magnification distributions were compared, with
the difference  critical convolution radius  defined as being  that at
which
\begin{equation}
\mbox{Max } \left( 2 \frac{\left| D^1_i - D^2_i \right|} {\left( D^1_i
+ D^2_i \right)} \right) < f
\end{equation}
where $D^1$  and $D^2$  are the cumulative  probability distributions;
such  cumulative distributions  ease differences  in  the differential
magnification probability distributions due  to noise.  The factor $f$
represents  the fractional  difference between  the  two magnification
distribution and for the purposes of this study we choose $f=0.05$ and
$f=0.10$ as representative values. Of  course, the choice of the value
of $f$  is rather arbitrary and  other choices could  be made.  Hence,
the  functional form employed  here should  be viewed  as illustrative
rather than  definitive.  The results of this  procedure are presented
in  Figure~\ref{fig3}; here  the  solid line  represents the  Einstein
radius of  the smaller  mass component, with  the points  denoting the
critical radius as defined by  the above criterion.  This figure shows
that  there is  a  clear trend  in  this statistic  with the  critical
Gaussian  radius increasing  in step  with the  Einstein  radius.  For
$f=0.05$   and   mass  scales   from   $\sim10^{-4}$M$_\odot$  up   to
$\sim$0.1M$_\odot$, the ratio of  these two quantities is (remarkably)
constant at $\sim4$  (with a couple of discrepant  points due to noise
spikes in the  cumulative distribution delaying convergence), although
this ratio  drops to $\sim1.1$  for $f=0.10$. Note, the  arrows denote
masses at which  smoothing on the scale of less  than 5 Einstein radii
failed  to  result  in   convergence  between  the  two  magnification
distributions.  A reexamination  of Figure~\ref{fig1} illustrates that
at this high mass end, the caustic structure introduced by the compact
masses is  similar in scale to  the overall caustic network  and it is
understandable that  this structure has  not been smoothed out  on the
scales under consideration.

Returning to Figure~\ref{fig2}  it is clear that the  third and fourth
row  of  panels  straddle  this critical  region,  with  magnification
distributions  for   the  smaller  Gaussian   source  appearing  quite
different for  the smooth matter  and compact cases.   Considering the
lower  panel,  and hence  larger  source,  the  distributions in  both
scenarios have  become very  similar. Notice in  this case,  even when
smoothed   with  the  larger   source,  the   resulting  magnification
distributions  possess  significant structure  which  is reflected  in
quite  dramatic events  in  the microlensing  light  curve. Hence  the
origin of the similarity in the distributions is not due to all of the
magnification structure  being smoothed out of  the microlensing maps.
For cases where the difference in mass between the compact dark matter
and the solar mass  stars is less, then a lot of  the structure in the
magnification maps  is smoothed out,  with the resulting  light curves
possessing very little structure.

\subsection{Further simulations}\label{further}
So far this paper has  considered a single combination of microlensing
optical depth  and shear,  but how general  are the  results uncovered
thus  far?    In  addressing  this   question,  three  more   sets  of
microlensing  parameters  were  employed,  namely  $(\sigma,\gamma)  =
(0.2,0.0)$,  $(\sigma,\gamma)  =  (0.2,0.2)$  and  $(\sigma,\gamma)  =
(0.6,0.6)$. As with the  previous simulations, the source region under
consideration covered 12$^2$ Einstein radii  for a solar mass star and
consisted of $2048^2$ pixels. Again, in each case, half of the optical
depth comprised  of solar mass stars  while the other half  was in the
form   of  smooth   matter  or   compact  objects,   of   mass  either
$3.5\times10^{-2}{\rm M_\odot}$ or $1.2\times10^{-3}{\rm M_\odot}$.

Rather   than    simply   repeating   the    analysis   presented   in
Section~\ref{analysis}, a slightly  different approach was undertaken.
As revealed in  Fig.~\ref{fig1}, as the masses of  the compact objects
decreases, the structure in the magnification map tends to that of the
smooth  matter case,  with the  smaller masses  providing  strong, but
localized,  perturbations, and  it  is these  perturbations which  are
washed out by convolving with a large enough source; note, however, if
the masses of the compact objects are large enough, they produce gross
changes into  the magnification map  such that convolving  cannot make
the magnification  identical to the smooth matter  case.  Hence, given
the  small masses  considered in  this second  set of  simulations, we
would  expect that  the structures  in the  compact matter  and smooth
matter magnification  maps should become similar, if  convolved with a
large enough source.

\begin{figure}
\centerline{ \psfig{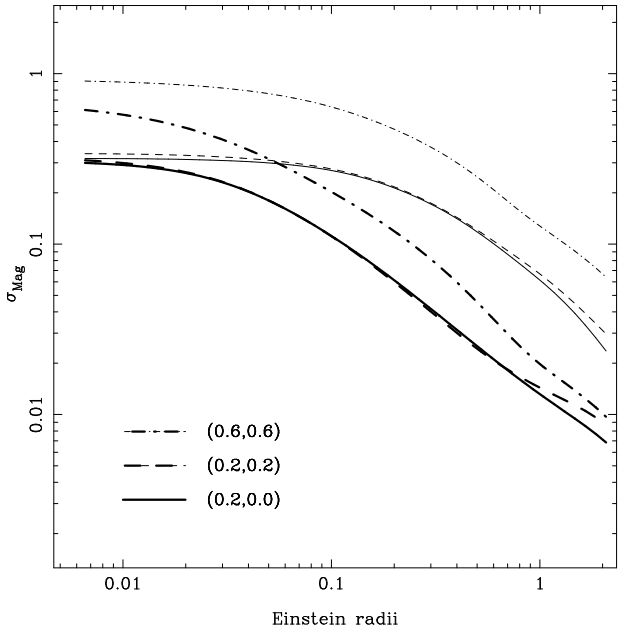}}
\caption[]{The  variation in  the  subtracted convolved  magnification
maps, as discussed in Section~\ref{further}. The differing line styles
correspond to  the three  parameter sets employed  (as denoted  by the
key)  where as  the  lighter  lines represent  the  compact masses  of
$3.5\times10^{-2}{\rm  M_\odot}$, whereas  the heavier  lines  are for
masses of  $1.2\times10^{-3}{\rm M_\odot}$.  Note that  the cases with
$\sigma=0.2$ possess similar structure and overlap in this figure.
\label{fig4}}
\end{figure}

The analysis procedure again consisted of convolving the smooth matter
and  compact matter  magnification maps  with a  Gaussian  profile.  A
residual map was constructed  by converting the smoothed magnification
maps to magnitudes and  then subtracting them. These residuals possess
a  Gaussian-like  profile,  centred  upon  zero,  and  so  a  residual
dispersion,  $\sigma_{Mag}$  was  calculated.   The  results  of  this
procedure are  presented in Figure~\ref{fig4}, in  which this residual
dispersion  is   plotted  against   the  smoothing  radius,   for  the
simulations under  consideration. The trend  is as expected,  with the
residual  map showing  large variations,  with  $\sigma_{Mag}\sim1$ at
small  smoothing  radii for  $(\sigma,\gamma)=(0.6,0.6)$,  but as  the
smoothing  radius is  increased, $\sigma_{Mag}$  falls, and  hence the
convolved   magnification  maps   present  similar   structure.  Also,
$\sigma_{Mag}$ falls  faster for the smaller compact  masses, as their
magnification structure gets washed out at smaller radii.

Interestingly,  the curves  for  the larger  mass  compact objects  in
Fig.~\ref{fig4}  clearly  show  a  plateau-like  structure,  remaining
constant  over a  range  of  smoothing radii,  and  then beginning  to
decrease. Furthermore,  the similarity of  the curves for  both masses
considered  suggests  that  the  smaller mass  component  possesses  a
similar plateau  structure, existing in  the regime below  the minimum
smoothing radius  we considered. Hence, this again  reveals that there
appears  to be  a critical  radius below  which the  influence  of the
compact  matter changes character,  effectively representing  a smooth
mass component. A systematic study of this critical radius for a range
of  microlensing parameters, and  the influence  of physical  size and
structure of the emission regions in quasars, will be the subject of a
forthcoming contribution.

\section{Implications}\label{implications}
One immediate  implication of the  results presented in this  paper is
that if  the dark matter  is in the  form of relatively  small compact
masses, then  the resulting  statistical properties of  a microlensing
light  curve will  be  dependent upon  the  size of  the source  under
consideration. It is important to note that this is somewhat different
to the usual size dependence of microlensing statistics where there is
a straightforward smoothing  of the light curve as  the source size is
increased, with a resultant narrowing of the magnification probability
distribution. Rather, a small  source and large source, where relative
sizes  of small  and large  can  be determined  from the  relationship
presented  in  Figure~\ref{fig3},  will  be subject  to  significantly
different magnification patterns and corresponding statistics.

The  effect influences  the question  of how  well we  can  reveal the
nature  of  the  dark  matter  component, if  present.   In  realistic
gravitational  microlensing scenarios, the  distribution of  matter in
the lensing galaxy will not  be a clean two-mass component population,
but  rather a certain  mass function  (or even  a combination  of mass
functions). Depending  upon the source  size for a given  system, this
mass function will suffer a cutoff point below which observations will
not be able  to distinguish between smooth and  granular matter.  This
does not  mean that  small lens masses  and large  sources combination
with  a single  mass component  cannot produce  microlensing imprints.
Similarly means that for large enough source sizes, sub-Jupiter masses
cannot be  detected unambiguously and that  claims involving planetary
and sub-planetary  populations that produce  light curves fluctuations
should  be  thoroughly justified  \citep[i.e.][]{2003ApJ...594...97C}.
For  example, from  Fig.~\ref{fig3},  for $f=0.05$,  we  see that  for
microlenses  masses of  $\sim$0.04~M$_\odot$, there  are  still source
sizes ($\sim$0.7~ER) that make such a low-mass population to appear as
a smooth distributed  matter component.  Considering the corresponding
physical   length  scale   in  microlensed   quasars   corresponds  to
$\sim0.04$pc \citep{1998ApJ...501..478L}, it is too large to represent
the  optical/UV  emitting  region   of  quasars.   However,  for  more
realistic  quasar  accretion disk  sizes  of $<$0.1~ER,  ``secondary''
populations  with  masses  $<$10$^{-3}$~M$_\odot$ will  be  completely
smoothed out.   Since the trend  in Fig.~\ref{fig3} is  quite constant
for both considered values of  $f$, this approach would also be useful
in putting  limits to source  sizes, though a better  understanding of
the role of the mass populations is needed.

To some  extent, it seems that  our results are  in contradiction with
those obtained by  Schechter et al. (2004). In  that work, the authors
concluded  that ``the magnification  probability distribution  for two
disparate components is not  \emph{exactly} that of a single component
and a smooth component''.  The apparent contradiction vanishes when it
is  realised that  Schechter et  al. (2004)  did not  account  for the
source size  effect (or, in other  words, they assumed  the same pixel
size  source in  their study).   Indeed, this  is an  important issue,
because  observationally the  source  effect will  be always  present.
This  means that the  additional structure  in the  magnification maps
induced by the low-mass component, seen by Schechter et al. (2004), is
again  blurred  out   by  the  source.  So,  the   dependence  of  the
magnification  probability on  the  higher-order moments  of the  mass
distributions is no longer true when introducing the source effect (at
least for certain regions of the mass distribution function).

\section{Conclusions}\label{conclusions}
This paper  has presented a study  of the influence of  source size on
the   properties  of   gravitational  microlensing   with   a  bimodal
distribution of lensing  masses.  Given the computational limitations,
the  study  focused  upon   a  set  of  macrolensing  parameters,  but
considered the  influence of  compact dark matter  on a range  of mass
scales. It was  found that small sources resolved  the caustic network
produced by the  smaller masses, but as the  source size was increased
this  fine scale  network becomes  washed out.   When compared  to the
scenario  where the  smaller masses  are replaced  with  an equivalent
quantity of  smooth matter, it  is seen that  the light curves  of the
smallest sources  in both circumstances  are quite different.   As the
source  size  is  increased,   the  light  curves,  and  corresponding
magnification   probability  distributions,   in   each  case   become
identical,  indicating that  there is  a critical  source  scale where
compact dark  matter behaves, in  a gravitational lensing  sense, like
smoothly distributed  matter. This critical source  scale is dependent
upon the ratio of the stellar to dark matter masses, with a relatively
linear  trend in  log-log space  (i.e. source  sizes smaller  than the
circles in  Figure~\ref{fig3} are small enough to  resolve the compact
dark  matter,  and  we  can  expect  to see  their  signature  in  the
microlensing  light curve, whereas  for source  size greater  than the
circles, this signature will be smoothed out).

The results  found in this  paper point to another  problem concerning
magnification    probabilities.    To   what    extent   magnification
distributions   are  showing   valuable  information   in  \emph{real}
gravitational  lensed systems?   In other  words, can  we characterize
systems  according to  their magnification  distributions?   To answer
these questions we would need to explore a large range of combinations
of $\sigma$  and $\gamma$,  together with different  mass distribution
functions; while computationally  expensive, we are currently planning
the first stages of such an exploration.  However, it seems clear that
when  the  source  effect  is  taken  into  account,  the  information
available from the magnification distributions can be rather poor.  Of
course,  the magnification  distributions  are only  the zeroth  level
statistic that can  be used to examine the  influence of gravitational
microlensing, but  clearly the results presented  in this contribution
similarly influence higher order  temporal statistics that are applied
to microlensing light curves.

In  closing,  this  contribution   returns  to  the  question  of  the
microlensing conjecture  that was discussed  earlier. The fall  of the
conjecture implies that  the mass function of the  compact objects was
potentially  amenable  to  microlensing  observations  as  it  is  now
apparent  that this  will imprint  a signature  into  any microlensing
light curves.   However, the issue  has been clouded further  with the
results presented in this paper illustrating that the size of a source
being of vital importance when addressing the question of how well the
properties of  an underlying mass  function can be  determined.  While
introducing  a mass  function  expands the  potential parameter  space
enormously, it is important to  understand the influence of the source
on   the  values   of   microlensing  statistics   and  hence   future
contributions will  investigate its role in  more general microlensing
scenarios.

\section*{Acknowledgments}
GFL thanks the Selby Trustees  for the 2004 Selby Research Award which
part  funded  this  project.  The  anonymous referee  is  thanked  for
comments which improved the paper.

\end{document}